\documentclass[twocolumn,times]{aastex63}
\usepackage{hyperref}
\usepackage{mleftright}

\accepted{October 6, 2020} 
\submitjournal{ApJL}

\shorttitle{Orbital Flash Foregrounds}
\shortauthors{Corbett et al.}

\begin{document}

\title{Orbital Foregrounds for Ultra-Short Duration Transients}

\correspondingauthor{Hank Corbett}
\email{htc@unc.edu}

\author[0000-0002-6339-6706]{Hank Corbett}
\affil{University of North Carolina at Chapel Hill, 120 E. Cameron Ave., Chapel Hill, NC 27514, USA}

\author[0000-0001-9380-6457]{Nicholas M. Law}
\affil{University of North Carolina at Chapel Hill, 120 E. Cameron Ave., Chapel Hill, NC 27514, USA}

\author[0000-0001-9981-4909]{Alan Vasquez Soto}
\affil{University of North Carolina at Chapel Hill, 120 E. Cameron Ave., Chapel Hill, NC 27514, USA}

\author[0000-0002-0583-0949]{Ward S. Howard}
\affil{University of North Carolina at Chapel Hill, 120 E. Cameron Ave., Chapel Hill, NC 27514, USA}

\author[0000-0001-9981-4909]{Amy Glazier}
\affil{University of North Carolina at Chapel Hill, 120 E. Cameron Ave., Chapel Hill, NC 27514, USA}

\author[0000-0001-5083-8272 ]{Ramses Gonzalez}
\affil{University of North Carolina at Chapel Hill, 120 E. Cameron Ave., Chapel Hill, NC 27514, USA}

\author[0000-0001-8791-7388]{Jeffrey K. Ratzloff}
\affil{University of North Carolina at Chapel Hill, 120 E. Cameron Ave., Chapel Hill, NC 27514, USA}

\author[0000-0001-8105-1042]{Nathan Galliher}
\affil{University of North Carolina at Chapel Hill, 120 E. Cameron Ave., Chapel Hill, NC 27514, USA}

\author[0000-0002-4227-9308]{Octavi Fors}
\affil{Institut de Ci\`encies del Cosmos (ICCUB), Universitat de Barcelona, IEEC-UB, Mart\'i i Franqu\`es 1, E08028 Barcelona, Spain}
\affil{University of North Carolina at Chapel Hill, 120 E. Cameron Ave., Chapel Hill, NC 27514, USA}

\author[0000-0001-9171-5236]{Robert Quimby}
\affil{San Diego State University, 5500 Campanile Dr., San Diego, CA 92182, USA}
\affil{Kavli Institute for the Physics and Mathematics of the Universe (WPI), The University of Tokyo Institutes for Advanced Study, The University of Tokyo, Kashiwa, Chiba 277-8583, Japan}

\begin{abstract}
Reflections from objects in Earth orbit can produce sub-second, star-like optical flashes
similar to astrophysical transients. Reflections have historically caused false alarms for transient
surveys, but the population has not been systematically studied. We report event rates for these
orbital flashes using the Evryscope Fast Transient Engine, a low-latency transient detection
pipeline for the Evryscopes. We select single-epoch detections likely caused by Earth satellites and
model the event rate as a function of both magnitude and sky position. We measure a rate of
$1800^{+600}_{-280}$ sky$^{-1}$ hour$^{-1}$, peaking at $m_g = 13.0$, for flashes morphologically
degenerate with real astrophysical signals in surveys like the Evryscopes. Of these,
$340^{+150}_{-85}$ sky$^{-1}$ hour$^{-1}$ are bright enough to be visible to the naked eye in
typical suburban skies with a visual limiting magnitude of $V\approx4$. These measurements place the
event rate of orbital flashes orders of magnitude higher than the combined rate of public alerts
from all active all-sky fast-timescale transient searches, including neutrino, gravitational-wave,
gamma-ray, and radio observatories. Short-timescale orbital flashes form a dominating foreground for
un-triggered searches for fast transients in low-resolution, wide-angle surveys. However, events
like fast radio bursts (FRBs) with arcminute-scale localization have a low probability
($\sim10^{-5}$) of coincidence with an orbital flash, allowing optical surveys to place constraints
on their potential optical counterparts in single images. Upcoming satellite internet
constellations, like SpaceX Starlink, are unlikely to contribute significantly to the population of
orbital flashes in normal operations.
\end{abstract}


\section{Introduction} \label{sec:intro} 

\begin{figure*}
    \centering
    \includegraphics[width=\textwidth]{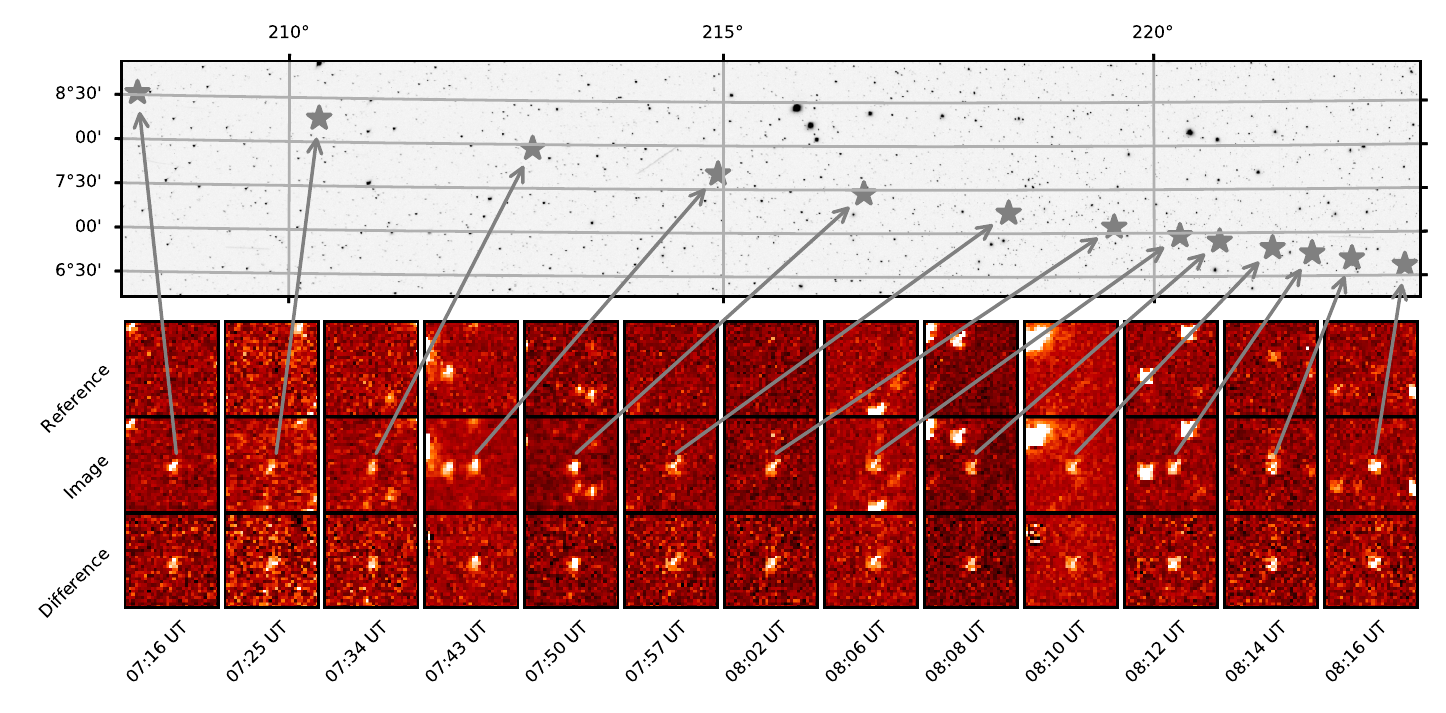}
    \caption{\label{fig:flash_trajectory} \textbf{(top)}~Example of a typical flash-producing
    trajectory seen by \textsc{EFTE}, followed over a single Evryscope pointing.
    \textbf{(bottom)}~Postage stamp cutouts of the reference, science, and discovery images,
    demonstrating point-like morphology. Each cutout is 30$\times$30 pixels (6.6$\times$6.6
    arcminutes) in size.}
\end{figure*}

Astrophysical phenomena with ultra-short (sub-minute) durations have largely escaped the scrutiny of
modern synoptic sky surveys, which are typically optimized for supernovae-like transients evolving
on day-to-month timescales. Sensitivity to minute- and hour-timescale events can be achieved through
sub-surveys over fractional sky areas; however, these searches typically use many-second exposures,
confining the shortest events to single images.

Timescale limitations are less of a factor for multi-messenger observatories, many of which operate
with nearly continuous 4$\pi$ sky coverage \citep{icecube_instrument, fermi_gbm_instrument,
ligo_instrument}. In the past decade, optical surveys with nightly sky coverage approaching that of
multi-messenger facilities have come online, including the Evryscopes
\citep{evryscope_project_paper, evryscope_instrument_paper}, the Mobile Astronomical System of
Telescope-Robots \citep[MASTER;][]{master_instrument},  the Asteroid Terrestrial-impact Last Alert
System \citep[ATLAS;][]{atlas_instrument}, the All-Sky Automated Survey for Supernovae
\citep[ASAS-SN;][]{asassn_instrument}, the Zwicky Transient Facility \citep[ZTF;][]{ztf_instrument}
and the Multi-site All-Sky CAmeRA \citep[MASCARA;][]{talens_2017}, opening up the rapid time domain
considerably; however, any single-image transients in these surveys are hidden in a fog of known
false positives, including particle strikes \citep{groom_2004} and reflected light from Earth
satellites, which can exhibit a broad range of morphologies.

Image contamination by Earth satellites takes two forms: streaks, with uniform illumination over
extended trajectories, and glints, which appear as short-duration flashes. These two morphologies
are frequently degenerate, and depend on the structure and orbit of the reflector. Streaks are
associated with fast-moving or slowly-rotating satellites, such as the Starlink constellation
discussed in \cite{mcdowell_2020}. Glints are associated with short rotation periods or high
altitudes, and are produced by chance alignments between an observer, the sun, and a reflective
rotating surface. The duration of a glint is the crossing time of the reflective surface's normal
vector across the disk of the sun, less than a second for satellites with minutes-long rotation
periods \citep{schaefer_1987}. Short durations relative to their motion on the sky and sharp
contrast with their associated streaks have led to glints being mistaken for astrophysical events
\citep{schaefer_1987,Maley_1987,Maley_1991, Rast_1991, shamir_2006}.

\cite{karpov_2016} presents time-resolved observations of satellite glints that reveal a peak in the
duration distribution at 0.4 seconds. Approximately half of the glints reported in
\cite{karpov_2016} were not coincident with the position of a satellite in the NORAD database.
Similarly, \cite{tingay_2020} noted multiple candidates with no or poorly-constrained association
with tracked satellites based on their latest two-line element parameters. Based on these
observations, it is unlikely that glints will be universally separable from a population of
astrophysical transients based on ephemerides for tracked satellites.  

The event rate and sky distribution of glints has not received systematic study, and the potential
for these events to contaminate searches for ultra-short transients remains. In this Letter, we use
single-image detections from the Evryscope Fast Transient Engine (Section~\ref{sec:observations}) to
measure the on-sky event rates of satellite glints for the first time. We provide estimates of the
flash rate as a function of observed magnitude and sky position (Section~\ref{sec:event_rates}). We
discuss the impact of this population on both current and upcoming observatory facilities, and its
implications for searches for ultra-short and multi-messenger transients, including the hypothesized
optical counterparts to Fast Radio Bursts (FRBs), in Section~\ref{sec:implications}.

\section{Observations and Survey} \label{sec:observations}

We used observations from the Evryscopes, a north-south pair of robotic all-sky telescopes, located
at Cerro Tololo Inter-American Observatory (Chile) and Mount Laguna Observatory (California). The
Evryscopes survey an instantaneous field of view (FOV) of 16,512 sq. degrees at 2-minute cadence
with 13.2\arcsec~{pixel}$^{-1}$ resolution. Images were collected between 2019 November 24
and 2020 April 16, and we have made no cuts for weather. Images within 10 degrees of the moon are
excluded. The Evryscope system design and survey strategy are described in
\cite{evryscope_instrument_paper} and \cite{evryscope_project_paper}.

The Evryscope Fast Transient Engine (\textsc{EFTE}) is a detection pipeline developed for
low-latency discovery of fast astrophysical transients, including bright flares from cool dwarfs
\citep{proxima_superflare, evryflare_i}, early phases of optical counterparts to gravitational wave
events \citep{corbett_gcn}, and the hypothesized optical counterparts to FRBs
\citep{lyutikov_2016,yang_2019, chen_2020}. 

\subsection{Single-Epoch Flash Sample}

We observed 3,372,044 high-probability candidates that passed our vetting criteria as described in
Section~\ref{sec:efte_pipeline}. Of these, we identify 1,415,722 candidates that do not appear in
multiple epochs as likely satellite flashes. This cut removes variable stars and persistent
artifacts from bright stars. 

Single-epoch candidates tend to occur in tracks across the sky on timescales ranging from
sub-cadence to hours. Figure~\ref{fig:flash_trajectory} shows a typical track observed by Evryscope,
followed for 1 hour, along with typical 30$\times$30 pixel (6.6$\times$6.6 arcminute)
subtraction stamps from the \textsc{EFTE} pipeline. The timescale of the delay between flashes gives
an angular speed of 10\arcsec~{second}$^{-1}$, or roughly one Evryscope pixel per second. However,
we do not observe any streaking at any epoch, implying that the duration of each individual flash is
much less than 1 second. This is consistent with the population of fast optical flashes noted in
\cite{biryukov_2015} and \cite{karpov_2016}, but observed in images integrated over minutes.

\subsection{Transient Detection with the Evryscopes} \label{sec:efte_pipeline}

\textsc{EFTE} uses a simplified image-subtraction technique for candidate detection. Evryscope focus
and optics, and thus PSFs, are stable on month-to-year timescales \citep{ratzloff_robotilter}, and
atmospheric seeing is dominated by optical effects under all observing conditions in each
Evryscope's 13.2\arcsec pixels. This stability means that image subtraction within a single
pointing does not require the PSF-matching techniques addressed by standard routines, such as
\textsc{HOTPANTS} \citep{hotpants} or \textsc{ZOGY} \citep{zogy}. Similarly, because we are
targeting the fastest-timescale events, we do not require widely-spaced reference frames; instead,
an earlier image from the same pointing is subtracted from each image, typically with a ten-minute
separation.

For each single-camera science image $S(x,y)$ and reference image $R(x,y)$, we calculate a discovery
image $D(x,y)$ from the per-pixel change in signal-to-noise ratio. Both $S(x,y)$ and $R(x, y)$ are
calibrated, background-subtracted, and aligned images in electron units. We measured image
background levels using an interpolated and clipped mesh, as implemented in \textsc{sextractor}
\citep{bertin_1996}. The discovery image $D(x,y)$ is given by:
\begin{equation}
	D(x,y) = \frac{S(x,y)}{s_{S}(x,y)} - \frac{R(x,y)}{s_{R}(x,y)},
\end{equation}
where $s_{R}(x,y)$ and $s_{S}(x,y)$ are noise images calculated for each image $I(x,y)$ with
measured background noise $s_B^2(x,y)$ as 
\begin{equation}
	s_I(x,y) = \sqrt{I(x,y) + s_B^2(x,y)}.
\end{equation}
While this technique is not statistically optimal, it is efficient (98.5\% of images are reduced in
cadence) and produces stable artifacts that can be rejected via automated means. We identified all
sources where $D(x,y) \geq 3$ in at least three contiguous pixels for automated vetting. We perform
aperture photometry for all candidates using the science image, and calibrated the results to the
ATLAS All-Sky Stellar Reference Catalog \citep[][]{atlas_catalog}.

A typical single image subtraction using this method produces $O(10^3)$ candidates. We require the
ratio of negative-to-positive pixels within a 5-pixel diameter to be less than 0.3 to remove dipole
artifacts from misalignments and Poisson noise peaks near bright stars, which reduces the number of
candidates to $O(10^2)$. These are then processed with a convolutional neural network
\citep{lecun_convnets} trained on small cutouts from $S(x,y)$, $D(x,y)$, and $R(x, y)$ in both real
and simulated data, reducing the final number of candidates to $6^{+13}_{-5}$ per image when
averaged over all weather conditions and cameras.

\subsection{Survey Completeness} \label{sec:completeness}

We characterized the completeness of the \textsc{EFTE} survey with injection-recovery testing. The
image sample contained 500 pairs with the same ten-minute separation used on-sky, and was
representative of all weather and instrument conditions contained in the survey. We injected
$\sim$1200 sources into each image ($8.0 \leq m_{g'} \leq 16.75$), using the normalized median of
nearby stars as the injection PSF.  We processed each image pair as described in
Section~\ref{sec:efte_pipeline} twice, alternating which image in the pair was used as the science
frame. We define a recovery as a candidate detected within three pixels of an injection position,
with the radius determined by the 99th percentile of the nearest-neighbor distance between the
vetted candidates and their nearest injection position.

\begin{figure}
    \centering
    \includegraphics{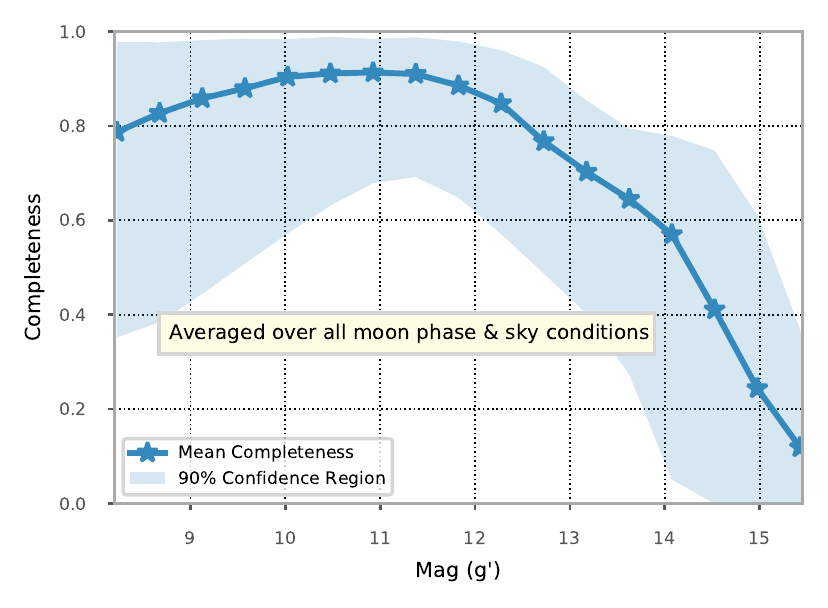}
    \caption{\label{fig:completeness_contamination} Completeness vs. magnitude for the Evryscope
    Fast Transient Engine (\textsc{EFTE}) pipeline, averaged over all observing conditions and
    cameras.}
\end{figure}
Figure~\ref{fig:completeness_contamination} shows the measured completeness as a function of
magnitude, along with the 90\% confidence interval (CI), which includes measurement uncertainty,
variability with observing conditions, and the vignetting effects. Completeness decreases at both
the bright and dim ends of the distribution. Saturation causes poorly-constrained centroids leading
to non-recovery at the bright end. The average 50\% completeness limit of $m_{g'} = 14.2$ is
brighter than the Evryscope dark-sky, single-image limit of $m_{g'} = 16$ due to a combination of
including all sky conditions in our analysis and using reference images with similar noise profiles
to science images.

\subsection{Candidate Reliability} \label{sec:reliability} 

Despite careful vetting, the median \textsc{EFTE} false positive rate (FPR) is $\sim$6,800 per hour.
In \textsc{EFTE} surveys for rapidly-evolving transients, further vetting is performed by
crossmatching with galaxy and stellar catalogs, or by association with gamma-ray and
gravitational-wave transient skymaps, combined with direct human interaction. Measuring an all-sky
event rate, however, requires knowing the fraction of real events in the sample. To measure this
fraction, we visually inspected 27,817 randomly selected single-epoch candidates that passed
our automated vetting as described in Section~\ref{sec:efte_pipeline}. 

Each candidate was assigned a binary classification, where a ``real'' classification corresponds to
a candidate that is morphologically consistent with an astrophysical transient. Of the candidates
classified, 2,823 were classified as real. From this, we estimate the real flash fraction (RFF) in
the vetted EFTE event stream to be $0.10\pm0.03$. Least-squares fits the to RFF as a function of
magnitude and solar elongation produced slopes not different from zero, with and uncertainty set by
the standard deviation of the residuals around the fit. Bogus candidates included:
\begin{enumerate}
    \setlength{\itemsep}{0pt}
    \setlength{\parskip}{0pt}
    \item Subtraction artifacts from bright stars ($50\%$).
    \item Optical ghosts, distinguished based on non-PSF-like shapes and presence in the reference
    frame ($1\%$).
    \item Aircraft strobes, distinguished based on nearby parallel streaks ($3\%$).
    \item Particle strikes ($46\%$). 
\end{enumerate}
The final category includes both readily-identifiable cosmic-ray muon tracks and signals caused by
Compton recoil electrons from environmental radioisotopes, which can be PSF-like. To constrain this
population, we reduced a series of 120-second darks with \textsc{EFTE} and searched for candidates
meeting our vetting criteria. We place an upper limit on the base rate of PSF-like particle strikes
of $\leq0.1$ per image, or $\sim300$ sky$^{-1}$ hour$^{-1}$, which is small compared to the all-sky
orbital flash rate (Section~\ref{sec:magnitude_rates}).

Evryscope PSFs, and the resulting degree to which stellar sources are undersampled, are
variable across each individual camera’s FOV. At the center of the field, dim sources can have a
FWHM $\ll$ 1 pixel. To avoid distortion of these sources, we estimate the prevalence of
particle strikes in our candidate sample using the RFF, rather than directly
removing them from the images with standard cosmic-ray mitigation tools, such as LA-Cosmic
\citep{lacosmic}.

\section{Event Rates of Flash Events} \label{sec:event_rates} 

We calculated event rates for candidates within discrete bins in observed magnitude in a two-minute
integration. Raw event rates for each magnitude bin $r(m_o)$ are given by:
\begin{equation}
    r(m_o) = \frac{F_m}{c_if_iN_i},
\end{equation}
where $F_m$ and $N_i$ are the number of candidates observed within the magnitude bin and the number
of images in the survey respectively, $c_i$ is the coverage per image, and $f_i$ is the fill
fraction of each image (\emph{i.e.}, the fraction of a single camera FOV that contributes to
the overlap-deduplicated Evryscope FOV). Per-image coverage is constant at $c_i = 12.348$~deg$^2$
hours. The fill fraction is determined by camera arrangement ($f_i = 0.965$). 

\subsection{Monte Carlo Rate Correction} \label{sec:rate_correction}

We used a Monte Carlo approach to model the effects of completeness and reliability as a function of
magnitude (Section~\ref{sec:completeness}). We modeled the per-magnitude completeness as a bounded
Johnson distribution \citep{johnson_1949} using maximum likelihood estimates for the mean, standard
deviation, skewness, and kurtosis from injection testing. We corrected for the reliability of the
event sample using the RFF described in Section~\ref{sec:reliability}, modeled per magnitude bin as
a normal distribution. 

For each magnitude bin, we made 100,000 draws from the fitted completeness and reliability
distributions, each time calculating a corrected event rate as
\begin{equation}
    \Gamma(m_o)_i = \frac{r_{m_o} R_i}{C_i},
\end{equation}
where $C$ and $R$ represent values drawn from the completeness and reliability distributions,
respectively. The reported mean rate is the 50th percentile of $\Gamma(m_o)_i$. The lower and upper
bounds of the 90\% CI were taken to be the 5th and 95th percentile of $\Gamma(m_o)_i$. 

\subsection{Magnitude Distribution} \label{sec:magnitude_rates}

As noted in \cite{lyutikov_2016}, flashes shorter than an image exposure time (typically $\gg1$
second) will exhibit phase blurring, diluting their flux relative to the surrounding stars by the
ratio of the integration time to their intrinsic duration. For sub-second durations, consistent with
the example seen in Figure~\ref{fig:flash_trajectory} and in \cite{biryukov_2015}, the peak of the
flash light curve will be brighter than is observed in long integrations. In general, the peak
magnitude of the flash $m_P$ is given by 
\begin{equation}\label{eqn:peak_mag}
    m_P = -2.5\log_{10}
    \mleft(
    \frac{T_{exp}10^{-0.4m_o}}{\tau_f}
    \mright), 
\end{equation}
where $\tau_f$ is the equivalent width of the light curve and $T_{exp}$ is the exposure time. We
assume a flash duration of 0.4 seconds, based on the mode of the distribution presented in
\cite{karpov_2016}, when estimating the peak brightness from the observed magnitude.

\begin{figure}
    \centering
    \includegraphics[width=\columnwidth]{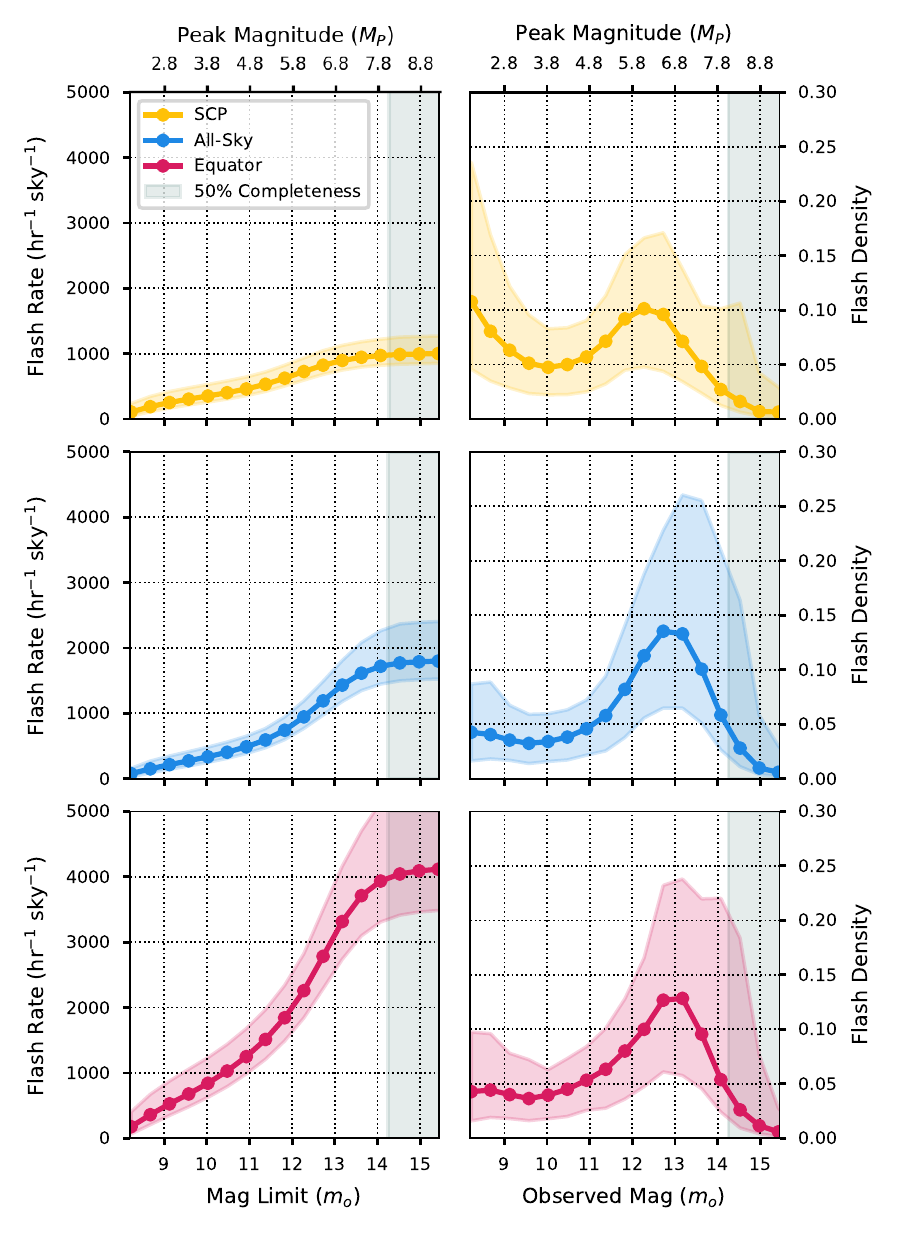}
    \caption{\label{fig:cumulative_event_rates} Cumulative \textbf{(left)} and normalized
    \textbf{(right)} flash rate distributions as a function of magnitude. We consider the event rate
    around the south celestial pole (SCP) \textbf{(top)}, averaged across the sky \textbf{(center)},
    and around the equator \textbf{(bottom)} as a function of both observed magnitude in 120 second
    integrations and limiting peak magnitudes assuming a 0.4 second flash duration. Rates are
    corrected using the technique described in Section~\ref{sec:rate_correction} to account for
    survey completeness and contamination. Shaded region gives the 90\% CI on each distribution.}
\end{figure}
Figure~\ref{fig:cumulative_event_rates} shows the cumulative and normalized magnitude density
distributions as a function of observed and estimated peak brightness. The shaded regions represent
the 90\% CI. For the cumulative distributions, the CI is bounded by the cumulative quadrature sum of
the per-bin CI limits. We extrapolate an all-sky event rate from the observed rates in three
regions: around the south celestial pole (SCP), within ten degrees of the equator, and across all
declinations. Both the all-sky and equatorial distributions peak at $m_o = 13$, whereas the polar
distribution peaks at $m_o = 12.2$, with a possible second peak near the saturation limit.

We measured integrated flash rates for $m_o < 14.25$ of $1.0^{+0.27}_{-0.15}\times 10^{3}$,
$4.0^{+1.40}_{-0.60}\times 10^{3}$, and $1.8^{+0.60}_{-0.28}\times 10^{3}$ sky$^{-1}$ hour$^{-1}$ in
the SCP, equatorial, and averaged all-sky regions respectively. Assuming that the observed
population is dominated by satellite glints, we expect a higher rate near the equator because
equatorial orbits are confined to a narrow declination band. The SCP region will only contain
objects in polar orbits spanning the full declination range, lowering the observed rate.

\subsection{Solar Geometry Dependence}\label{sec:solar_geometry}

\begin{figure}
    \centering
    \includegraphics[width=\columnwidth]{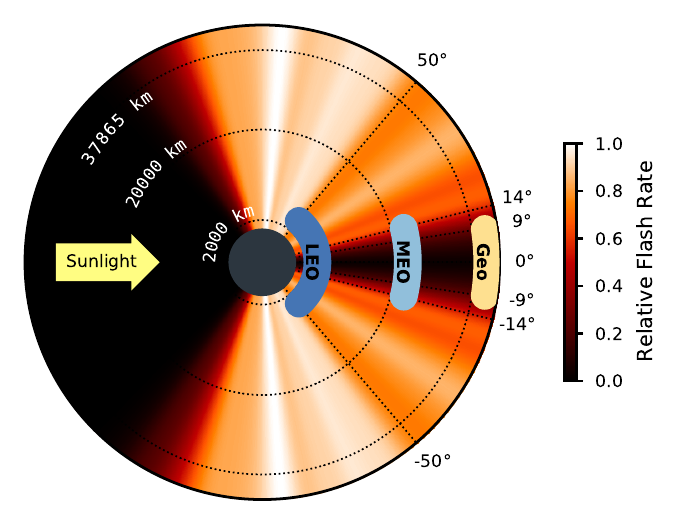}
    \caption{\label{fig:antipode_geometry} Geometry of Earth's shadow relative to the position of
    the solar antipode. The angular size of the shadow depends on altitude, ranging from 9 degrees
    for a typical geosynchronous orbit, to 50 degrees at the upper limit of low-Earth orbit. The
    color gradient represents a kernel density estimation for the relative flash rate as a function
    of antipode distance based on the visually-sorted sample from Section~\ref{sec:reliability}, and
    is reflected across the Sun-Earth axis.}
\end{figure}

If the observed flashes are caused by reflections from satellites, none should be expected from the
region of the sky covered by Earth's shadow. The solid angle subtended by the shadow depends on
satellite altitude, following the geometry illustrated in Figure~\ref{fig:antipode_geometry}. For
low-Earth orbit (LEO) altitudes ($<2,000$ km), the shadow covers a 50 degree radius around the solar
antipode. The angular size shrinks to a 14 degree radius for medium-Earth orbit (MEO), or a 9 degree
radius for geosynchronous orbit.
We evaluated the distance between each of the human-vetted candidates from
Section~\ref{sec:reliability} and the solar antipode. Shaded regions depict the angular extent of
Earth's shadow for LEO, MEO, and geosynchronous satellites.

The prevalence of flashes decreases steadily with proximity to the shadow in the region covered for
LEO objects, before falling in the solid angle covered for MEO and higher orbits. Approximately 34\%
of the flashes occurred within 50 degrees of the center of the shadow. Few flashes occur within the
region within Earth's shadow for MEO and geosynchronous orbits, with 3.5\% within 14 degrees of the
antipode and only 1.1\% within 9 degrees, which suggests that the majority of the flashes are
generated by satellites in middle- and high-Earth orbit. No Evryscope observations occur within
$\sim$80 degrees of the sun. 

\section{Implications for Fast Transient Searches} \label{sec:implications} 

Earth satellites produce thousands of potential false alarms mimicking fast transients. One option
to mitigate this fog is to duplicate monitoring across a substantial ($\sim3$ kilometer for
Evryscope-scale resolution) baseline, enabling parallax-based coincidence rejection. However, this
requires construction of multiple facilities. An alternative approach is to track glint-producing
objects directly in the event stream, using calculated orbit fits to reject candidates that occur in
tracks. This approach is currently in development for the \textsc{EFTE} event stream (Vasquez~Soto
et al. 2020, in prep). 

Degradation of the night sky by ``megaconstellations" of LEO satellites is anticipated to be a major
environmental challenge for astronomy in the coming decade \citep{mcdowell_2020}. The
construction of these constellations will increase the number of artificial satellites in orbit by a
factor of many, with a corresponding increase in the amount of reflected sunlight visible in
astronomical images. However, due to their high angular speeds and controlled rotation, reflected
light from satellites similar to Starlink is unlikely to produce PSF-like glints during normal
operations.

\subsection{Visual Observers}

The all-sky magnitude distribution in Figure~\ref{fig:cumulative_event_rates} shows that the
instantaneous peak brightness of many flashes should be detectable to the unassisted human eye. For
a typical suburban sky with limiting magnitude $V\approx4$, we predict a naked-eye-visible event
rate of $340^{+150}_{-85}$ sky$^{-1}$ hour$^{-1}$ based on the all-sky averaged cumulative
event rate of $1.8^{+0.60}_{-0.28}\times 10^{3}$ sky$^{-1}$ hour$^{-1}$ ($m_o < 14.25$). At the
darkest sky sites, with limiting magnitude $V\approx6$, this rate increases
to $740^{+200}_{-140}$ sky$^{-1}$ hour$^{-1}$. Within the equatorial region, the
expected naked-eye event rate increases to $840^{+390}_{-220}$ sky$^{-1}$ hour$^{-1}$ ($V\approx4$),
or $1800^{+500}_{-350}$ sky$^{-1}$ hour$^{-1}$ ($V\approx6$), based on the cumulative equatorial
event rate of $4.0^{+1.40}_{-0.60}\times 10^{3}$ sky$^{-1}$ hour$^{-1}$.

The time resolution \citep{cornsweet_2014} and biochemical adaptation time \citep{dunn_2006,
yeh_lee_kremers_1996} of the human eye are comparable to orbital flash durations. We expect the
naked-eye detectability of these flashes to vary strongly with observer ability. 

\subsection{Narrow-Field Imaging}

The impact of orbital flashes on narrow-field imagers is negligible. For a 10$\times$10 arcminute
FOV, we predict approximately 1.2 flashes per 1,000 hours of exposure time based on the
all-sky averaged event rate, or 2.7 flashes per 1,000 hours of exposure time based on the equatorial
region event rate. While these are relatively rare events, they could account for occasional
flashes that have been remarked on by amateur astronomers and some rejected single-image detections
in tiling sky surveys. 

\subsection{Multi-Messenger Coincidence Searches}

The flash rate measured here also implies a high coincidence rate for multi-messenger events. The
Canadian Hydrogen Intensity Mapping Experiment (CHIME) is able to localize FRBs to the nearest
arcminute \citep{chime_2018}, reducing the expected flash count within the error radius to a
likely-acceptable $3.8 \times 10^{-5}$ hour$^{-1}$ based on the all-sky averaged event rate.
Wide-angle surveys like Evryscope can be expected to have a false alarm rate (FAR) of 1 per 3 years
for apparent FRB optical counterparts due to orbital flashes. For FRBs localized to the
equatorial region, the expected event rate and FAR increase to $8.5 \times 10^{-5}$ hour$^{-1}$ and
1 per 1.3 years, respectively.

In contrast, gamma-ray bursts (GRBs) from the Fermi Gamma Burst Monitor (GBM) have a median 90\%
localization area of 209 sq. degrees \citep{goldstein_2020}, leading to an expected flash count of
9.1~hour$^{-1}$ based on the all-sky averaged cumulative event rate. With simultaneous
coverage and physical constraints on the timescale of the optical component, the expected flash
count drops to 0.15~minute$^{-1}$. For GRB localizations within the equatorial region, the
rates increase to 20~hour$^{-1}$, or 0.33~minute$^{-1}$.

Events with larger localization regions or weakly-constrained early lightcurves, like gravitational
wave events from LIGO/Virgo, will be more heavily impacted. For a typical 1200 sq. degree sky map,
the expected flash is 53~hour$^{-1}$ assuming the all-sky averaged event rate, or
120~hour$^{-1}$ within the equatorial region. The resulting FAR will increase linearly with both
localization area and trigger rate. 

\subsection{Vera C. Rubin Observatory} \label{sec:rubin}

 We predict that point-like flashes will occur in images from Vera C. Rubin Observatory at a
rate of $6.4 \times 10^{-4}$ flashes per image (15 seconds of 3.5 sq. degrees), based on the
all-sky averaged event rate. Assuming $O(1000)$ pointings, we expect Rubin to observe on the order
of 1.3 flashes per night from this population. In the worst case scenario of all pointings
being confined to the equatorial region, this rate increases to $1.4 \time 10^{-3}$ flashes per
image, or 2.8 flashes per night. Due to the 15-second exposure time of Rubin, the average observed
magnitude of the flash distribution will be shifted to $m_g\sim 10.7$, $100\times$ brighter than the
$g$-band bright limit of $m_g = 15.7$ \citep{LSST_Science_Book}. Our observed magnitude distribution
drops off at the faint end, but there is the potential for a second distribution to exist beyond the
Evryscope depth limit, as is seen for geosynchronous debris observed by the DebrisWatch survey
\citep{blake_2020}.

We estimate that a worst-case scenario 10 millisecond flash from a geosynchronous satellite will
produce a 0.15\arcsec~streak, likely indistinguishable in Rubin's 0.2\arcsec~pixels. Longer-duration
flashes, like the 0.4 second events we have assumed here, would produce obvious 6\arcsec~streaks.
However, fully constraining the expected morphology of orbital flashes in Rubin data will require
modeling of their duration via orbital characteristics of the population.

\section*{Acknowledgements}

The Evryscope was constructed under National Science Foundation ATI grant AST-1407589, with
operating costs from National Science Foundation CAREER grant 1555175. Current operations are
supported by AAG-2009645. HC was supported by the National Science Foundation Graduate Research
Fellowship (Grant No. DGE-1144081), AAG-2009645, and the North Carolina Space Grant. OF acknowledges
the support by the Spanish Ministerio de Ciencia e Innovaci\'{o}n (MICINN) under grant
PID2019-105510GB-C31 and through the ``Center of Excellence Mar\'{i}a de Maeztu 2020-2023'' award to
the ICCUB (CEX2019-000918-M). This research made use of
Astropy,\footnote{\href{http://www.astropy.org}{http://www.astropy.org}} a community-developed core
Python package for Astronomy \citep{astropy_2013, astropy_2018}, and
SciPy,\footnote{\href{http://www.scipy.org}{http://www.scipy.org}} a core Python package for general
scientific computing tasks. 


\begin{thebibliography}{}
    \expandafter\ifx\csname natexlab\endcsname\relax\def\natexlab#1{#1}\fi
    \providecommand{\url}[1]{\href{#1}{#1}}
    \providecommand{\dodoi}[1]{doi:~\href{http://doi.org/#1}{\nolinkurl{#1}}}
    \providecommand{\doeprint}[1]{\href{http://ascl.net/#1}{\nolinkurl{http://ascl.net/#1}}}
    \providecommand{\doarXiv}[1]{\href{https://arxiv.org/abs/#1}{\nolinkurl{https://arxiv.org/abs/#1}}}
    
    \bibitem[{{Abbott} {et~al.}(2009){Abbott}, {Abbott}, {Adhikari}, {Ajith},
      {Allen}, {Allen}, {Amin}, {Anderson}, {Anderson}, {Arain}, {Araya}, {Armand
      ula}, {Armor}, {Aso}, {Aston}, {Aufmuth}, {Aulbert}, {Babak}, {Baker},
      {Ballmer}, {Barker}, {Barker}, {Barr}, {Barriga}, {Barsotti}, {Barton},
      {Bartos}, {Bassiri}, {Bastarrika}, {Behnke}, {Benacquista}, {Betzwieser},
      {Beyersdorf}, {Bilenko}, {Billingsley}, {Biswas}, {Black}, {Blackburn},
      {Blackburn}, {Blair}, {Bland}, {Bodiya}, {Bogue}, {Bork}, {Boschi}, {Bose},
      {Brady}, {Braginsky}, {Brau}, {Bridges}, {Brinkmann}, {Brooks}, {Brown},
      {Brummit}, {Brunet}, {Bullington}, {Buonanno}, {Burmeister}, {Byer},
      {Cadonati}, {Camp}, {Cannizzo}, {Cannon}, {Cao}, {Cardenas}, {Caride},
      {Castaldi}, {Caudill}, {Cavagli{\`a}}, {Cepeda}, {Chalermsongsak},
      {Chalkley}, {Charlton}, {Chatterji}, {Chelkowski}, {Chen}, {Christensen},
      {Chung}, {Clark}, {Clark}, {Clayton}, {Cokelaer}, {Colacino}, {Conte},
      {Cook}, {Corbitt}, {Cornish}, {Coward}, {Coyne}, {Creighton}, {Creighton},
      {Cruise}, {Culter}, {Cumming}, {Cunningham}, {Danilishin}, {Danzmann},
      {Daudert}, {Davies}, {Daw}, {DeBra}, {Degallaix}, {Dergachev}, {Desai},
      {DeSalvo}, {Dhurandhar}, {D{\'\i}az}, {Dietz}, {Donovan}, {Dooley}, {Doomes},
      {Drever}, {Dueck}, {Duke}, {Dumas}, {Dwyer}, {Echols}, {Edgar}, {Effler},
      {Ehrens}, {Espinoza}, {Etzel}, {Evans}, {Evans}, {Fairhurst}, {Faltas},
      {Fan}, {Fazi}, {Fehrmenn}, {Finn}, {Flasch}, {Foley}, {Forrest},
      {Fotopoulos}, {Franzen}, {Frede}, {Frei}, {Frei}, {Freise}, {Frey}, {Fricke},
      {Fritschel}, {Frolov}, {Fyffe}, {Galdi}, {Garofoli}, {Gholami}, {Giaime},
      {Giampanis}, {Giardina}, {Goda}, {Goetz}, {Goggin}, {Gonz{\'a}lez},
      {Gorodetsky}, {Go{\ss}ler}, {Gouaty}, {Grant}, {Gras}, {Gray}, {Gray},
      {Greenhalgh}, {Gretarsson}, {Grimaldi}, {Grosso}, {Grote}, {Grunewald},
      {Guenther}, {Gustafson}, {Gustafson}, {Hage}, {Hallam}, {Hammer}, {Hammond},
      {Hanna}, {Hanson}, {Harms}, {Harry}, {Harry}, {Harstad}, {Haughian},
      {Hayama}, {Heefner}, {Heng}, {Heptonstall}, {Hewitson}, {Hild}, {Hirose},
      {Hoak}, {Hodge}, {Holt}, {Hosken}, {Hough}, {Hoyland}, {Hughey}, {Huttner},
      {Ingram}, {Isogai}, {Ito}, {Ivanov}, {Johnson}, {Johnson}, {Jones}, {Jones},
      {Jones}, {Ju}, {Kalmus}, {Kalogera}, {Kandhasamy}, {Kanner}, {Kasprzyk},
      {Katsavounidis}, {Kawabe}, {Kawamura}, {Kawazoe}, {Kells}, {Keppel},
      {Khalaidovski}, {Khalili}, {Khan}, {Khazanov}, {King}, {Kissel}, {Klimenko},
      {Kokeyama}, {Kondrashov}, {Kopparapu}, {Korand a}, {Kozak}, {Krishnan},
      {Kumar}, {Kwee}, {Lam}, {Landry}, {Lantz}, {Lazzarini}, {Lei}, {Lei},
      {Leindecker}, {Leonor}, {Li}, {Lin}, {Lindquist}, {Littenberg}, {Lockerbie},
      {Lodhia}, {Longo}, {Lormand}, {Lu}, {Lubi{\'n}ski}, {Lucianetti}, {L{\"u}ck},
      {Machenschalk}, {MacInnis}, {Mageswaran}, {Mailand}, {Mandel}, {Mandic},
      {M{\'a}rka}, {M{\'a}rka}, {Markosyan}, {Markowitz}, {Maros}, {Martin},
      {Martin}, {Marx}, {Mason}, {Matichard}, {Matone}, {Matzner}, {Mavalvala},
      {McCarthy}, {McClelland}, {McGuire}, {McHugh}, {McIntyre}, {McKechan},
      {McKenzie}, {Mehmet}, {Melatos}, {Melissinos}, {Men{\'e}ndez}, {Mendell},
      {Mercer}, {Meshkov}, {Messenger}, {Meyer}, {Miller}, {Minelli}, {Mino},
      {Mitrofanov}, {Mitselmakher}, {Mittleman}, {Miyakawa}, {Moe}, {Mohanty},
      {Mohapatra}, {Moreno}, {Morioka}, {Mors}, {Mossavi}, {Mow Lowry}, {Mueller},
      {M{\"u}ller-Ebhardt}, {Muhammad}, {Mukherjee}, {Mukhopadhyay}, {Mullavey},
      {Munch}, {Murray}, {Myers}, {Myers}, {Nash}, {Nelson}, {Newton}, {Nishizawa},
      {Numata}, {O'Dell}, {O'Reilly}, {O'Shaughnessy}, {Ochsner}, {Ogin},
      {Ottaway}, {Ottens}, {Overmier}, {Owen}, {Pan}, {Pankow}, {Papa},
      {Parameshwaraiah}, {Patel}, {Pedraza}, {Penn}, {Perraca}, {Pierro}, {Pinto},
      {Pitkin}, {Pletsch}, {Plissi}, {Postiglione}, {Principe}, {Prix},
      {Prokhorov}, {Punken}, {Quetschke}, {Raab}, {Rabeling}, {Radkins}, {Raffai},
      {Raics}, {Rainer}, {Rakhmanov}, {Raymond}, {Reed}, {Reed}, {Rehbein}, {Reid},
      {Reitze}, {Riesen}, {Riles}, {Rivera}, {Roberts}, {Robertson}, {Robinson},
      {Robinson}, {Roddy}, {R{\"o}ver}, {Rollins}, {Romano}, {Romie}, {Rowan},
      {R{\"u}diger}, {Russell}, {Ryan}, {Sakata}, {de la Jordana}, {Sandberg},
      {Sannibale}, {Santamar{\'\i}a}, {Saraf}, {Sarin}, {Sathyaprakash}, {Sato},
      {Satterthwaite}, {Saulson}, {Savage}, {Savov}, {Scanlan}, {Schilling},
      {Schnabel}, {Schofield}, {Schulz}, {Schutz}, {Schwinberg}, {Scott}, {Scott},
      {Searle}, {Sears}, {Seifert}, {Sellers}, {Sengupta}, {Sergeev}, {Shapiro},
      {Shawhan}, {Shoemaker}, {Sibley}, {Siemens}, {Sigg}, {Sinha}, {Sintes},
      {Slagmolen}, {Slutsky}, {Smith}, {Smith}, {Smith}, {Somiya}, {Sorazu},
      {Stein}, {Stein}, {Steplewski}, {Stochino}, {Stone}, {Strain}, {Strigin},
      {Stroeer}, {Stuver}, {Summerscales}, {Sun}, {Sung}, {Sutton}, {Szokoly},
      {Talukder}, {Tang}, {Tanner}, {Tarabrin}, {Taylor}, {Taylor}, {Thacker},
      {Thorne}, {Th{\"u}ring}, {Tokmakov}, {Torres}, {Torrie}, {Traylor}, {Trias},
      {Ugolini}, {Ulmen}, {Urbanek}, {Vahlbruch}, {Vallisneri}, {van den Broeck},
      {van der Sluys}, {van Veggel}, {Vass}, {Vaulin}, {Vecchio}, {Veitch},
      {Veitch}, {Veltkamp}, {Villar}, {Vorvick}, {Vyachanin}, {Waldman}, {Wallace},
      {Ward}, {Weidner}, {Weinert}, {Weinstein}, {Weiss}, {Wen}, {Wen}, {Wette},
      {Whelan}, {Whitcomb}, {Whiting}, {Wilkinson}, {Willems}, {Williams},
      {Williams}, {Willke}, {Wilmut}, {Winkelmann}, {Winkler}, {Wipf}, {Wiseman},
      {Woan}, {Wooley}, {Worden}, {Wu}, {Yakushin}, {Yamamoto}, {Yan}, {Yoshida},
      {Zanolin}, {Zhang}, {Zhang}, {Zhao}, {Zotov}, {Zucker}, {M{\"u}hlen}, \&
      {Zweizig}}]{ligo_instrument}
    {Abbott}, B.~P., {Abbott}, R., {Adhikari}, R., {et~al.} 2009, Reports on
      Progress in Physics, 72, 076901, \dodoi{10.1088/0034-4885/72/7/076901}
    
    \bibitem[{{Astropy Collaboration} {et~al.}(2013){Astropy Collaboration},
      {Robitaille}, {Tollerud}, {Greenfield}, {Droettboom}, {Bray}, {Aldcroft},
      {Davis}, {Ginsburg}, {Price-Whelan}, {Kerzendorf}, {Conley}, {Crighton},
      {Barbary}, {Muna}, {Ferguson}, {Grollier}, {Parikh}, {Nair}, {Unther},
      {Deil}, {Woillez}, {Conseil}, {Kramer}, {Turner}, {Singer}, {Fox}, {Weaver},
      {Zabalza}, {Edwards}, {Azalee Bostroem}, {Burke}, {Casey}, {Crawford},
      {Dencheva}, {Ely}, {Jenness}, {Labrie}, {Lim}, {Pierfederici}, {Pontzen},
      {Ptak}, {Refsdal}, {Servillat}, \& {Streicher}}]{astropy_2013}
    {Astropy Collaboration}, {Robitaille}, T.~P., {Tollerud}, E.~J., {et~al.} 2013,
      \aap, 558, A33, \dodoi{10.1051/0004-6361/201322068}
    
    \bibitem[{{Astropy Collaboration} {et~al.}(2018){Astropy Collaboration},
      {Price-Whelan}, {Sip{\H{o}}cz}, {G{\"u}nther}, {Lim}, {Crawford}, {Conseil},
      {Shupe}, {Craig}, {Dencheva}, {Ginsburg}, {Vand erPlas}, {Bradley},
      {P{\'e}rez-Su{\'a}rez}, {de Val-Borro}, {Aldcroft}, {Cruz}, {Robitaille},
      {Tollerud}, {Ardelean}, {Babej}, {Bach}, {Bachetti}, {Bakanov}, {Bamford},
      {Barentsen}, {Barmby}, {Baumbach}, {Berry}, {Biscani}, {Boquien}, {Bostroem},
      {Bouma}, {Brammer}, {Bray}, {Breytenbach}, {Buddelmeijer}, {Burke},
      {Calderone}, {Cano Rodr{\'\i}guez}, {Cara}, {Cardoso}, {Cheedella}, {Copin},
      {Corrales}, {Crichton}, {D'Avella}, {Deil}, {Depagne}, {Dietrich}, {Donath},
      {Droettboom}, {Earl}, {Erben}, {Fabbro}, {Ferreira}, {Finethy}, {Fox},
      {Garrison}, {Gibbons}, {Goldstein}, {Gommers}, {Greco}, {Greenfield},
      {Groener}, {Grollier}, {Hagen}, {Hirst}, {Homeier}, {Horton}, {Hosseinzadeh},
      {Hu}, {Hunkeler}, {Ivezi{\'c}}, {Jain}, {Jenness}, {Kanarek}, {Kendrew},
      {Kern}, {Kerzendorf}, {Khvalko}, {King}, {Kirkby}, {Kulkarni}, {Kumar},
      {Lee}, {Lenz}, {Littlefair}, {Ma}, {Macleod}, {Mastropietro}, {McCully},
      {Montagnac}, {Morris}, {Mueller}, {Mumford}, {Muna}, {Murphy}, {Nelson},
      {Nguyen}, {Ninan}, {N{\"o}the}, {Ogaz}, {Oh}, {Parejko}, {Parley}, {Pascual},
      {Patil}, {Patil}, {Plunkett}, {Prochaska}, {Rastogi}, {Reddy Janga},
      {Sabater}, {Sakurikar}, {Seifert}, {Sherbert}, {Sherwood-Taylor}, {Shih},
      {Sick}, {Silbiger}, {Singanamalla}, {Singer}, {Sladen}, {Sooley},
      {Sornarajah}, {Streicher}, {Teuben}, {Thomas}, {Tremblay}, {Turner},
      {Terr{\'o}n}, {van Kerkwijk}, {de la Vega}, {Watkins}, {Weaver}, {Whitmore},
      {Woillez}, {Zabalza}, \& {Astropy Contributors}}]{astropy_2018}
    {Astropy Collaboration}, {Price-Whelan}, A.~M., {Sip{\H{o}}cz}, B.~M., {et~al.}
      2018, \aj, 156, 123, \dodoi{10.3847/1538-3881/aabc4f}
    
    \bibitem[{{Becker}(2015)}]{hotpants}
    {Becker}, A. 2015, {HOTPANTS: High Order Transform of PSF ANd Template
      Subtraction}.
    \newblock \doeprint{1504.004}
    
    \bibitem[{{Bellm} {et~al.}(2019){Bellm}, {Kulkarni}, {Graham}, {Dekany},
      {Smith}, {Riddle}, {Masci}, {Helou}, {Prince}, {Adams}, {Barbarino},
      {Barlow}, {Bauer}, {Beck}, {Belicki}, {Biswas}, {Blagorodnova}, {Bodewits},
      {Bolin}, {Brinnel}, {Brooke}, {Bue}, {Bulla}, {Burruss}, {Cenko}, {Chang},
      {Connolly}, {Coughlin}, {Cromer}, {Cunningham}, {De}, {Delacroix}, {Desai},
      {Duev}, {Eadie}, {Farnham}, {Feeney}, {Feindt}, {Flynn}, {Franckowiak},
      {Frederick}, {Fremling}, {Gal-Yam}, {Gezari}, {Giomi}, {Goldstein},
      {Golkhou}, {Goobar}, {Groom}, {Hacopians}, {Hale}, {Henning}, {Ho}, {Hover},
      {Howell}, {Hung}, {Huppenkothen}, {Imel}, {Ip}, {Ivezi{\'c}}, {Jackson},
      {Jones}, {Juric}, {Kasliwal}, {Kaspi}, {Kaye}, {Kelley}, {Kowalski},
      {Kramer}, {Kupfer}, {Landry}, {Laher}, {Lee}, {Lin}, {Lin}, {Lunnan},
      {Giomi}, {Mahabal}, {Mao}, {Miller}, {Monkewitz}, {Murphy}, {Ngeow},
      {Nordin}, {Nugent}, {Ofek}, {Patterson}, {Penprase}, {Porter}, {Rauch},
      {Rebbapragada}, {Reiley}, {Rigault}, {Rodriguez}, {van Roestel}, {Rusholme},
      {van Santen}, {Schulze}, {Shupe}, {Singer}, {Soumagnac}, {Stein}, {Surace},
      {Sollerman}, {Szkody}, {Taddia}, {Terek}, {Van Sistine}, {van Velzen},
      {Vestrand}, {Walters}, {Ward}, {Ye}, {Yu}, {Yan}, \&
      {Zolkower}}]{ztf_instrument}
    {Bellm}, E.~C., {Kulkarni}, S.~R., {Graham}, M.~J., {et~al.} 2019, \pasp, 131,
      018002, \dodoi{10.1088/1538-3873/aaecbe}
    
    \bibitem[{{Bertin} \& {Arnouts}(1996)}]{bertin_1996}
    {Bertin}, E., \& {Arnouts}, S. 1996, \aaps, 117, 393,
      \dodoi{10.1051/aas:1996164}
    
    \bibitem[{{Bhat} {et~al.}(2009){Bhat}, {Meegan}, {Lichti}, {Briggs},
      {Connaughton}, {Diehl}, {Fishman}, {Greiner}, {Kippen}, {Kouveliotou},
      {Paciesas}, {Preece}, \& {von Kienlin}}]{fermi_gbm_instrument}
    {Bhat}, P.~N., {Meegan}, C.~A., {Lichti}, G.~G., {et~al.} 2009, in American
      Institute of Physics Conference Series, Vol. 1133, American Institute of
      Physics Conference Series, ed. C.~{Meegan}, C.~{Kouveliotou}, \&
      N.~{Gehrels}, 34--36, \dodoi{10.1063/1.3155916}
    
    \bibitem[{{Biryukov} {et~al.}(2015){Biryukov}, {Beskin}, {Karpov}, {Bondar},
      {Ivanov}, {Katkova}, {Perkov}, \& {Sasyuk}}]{biryukov_2015}
    {Biryukov}, A., {Beskin}, G., {Karpov}, S., {et~al.} 2015, Baltic Astronomy,
      24, 100, \dodoi{10.1515/astro-2017-0208}
    
    \bibitem[{{Blake} {et~al.}(2020){Blake}, {Chote}, {Pollacco}, {Feline},
      {Privett}, {Ash}, {Eves}, {Greenwood}, {Harwood}, {Marsh}, {Veras}, \&
      {Watson}}]{blake_2020}
    {Blake}, J.~A., {Chote}, P., {Pollacco}, D., {et~al.} 2020, arXiv e-prints,
      arXiv:2008.12799.
    \newblock \doarXiv{2008.12799}
    
    \bibitem[{{Chen} {et~al.}(2020){Chen}, {Ravi}, \& {Lu}}]{chen_2020}
    {Chen}, G., {Ravi}, V., \& {Lu}, W. 2020, \apj, 897, 146,
      \dodoi{10.3847/1538-4357/ab982b}
    
    \bibitem[{{Corbett} {et~al.}(2019){Corbett}, {Ackley}, {Law}, {Gonzalez
      Chavez}, {Vasquez}, {Ratzloff}, {Eikenberry}, {Howard}, {Glazier},
      {Galliher}, {Reichart}, {Haislip}, {Kouprianov}, \& {Quimby}}]{corbett_gcn}
    {Corbett}, H., {Ackley}, K., {Law}, N., {et~al.} 2019, GRB Coordinates Network,
      26227, 1
    
    \bibitem[{Cornsweet(2014)}]{cornsweet_2014}
    Cornsweet, T. 2014, Visual Perception (Elsevier Science)
    
    \bibitem[{Dunn \& Rieke(2006)}]{dunn_2006}
    Dunn, F.~A., \& Rieke, F. 2006, Current Opinion in Neurobiology, 16, 363–370,
      \dodoi{10.1016/j.conb.2006.06.013}
    
    \bibitem[{Goldstein {et~al.}(2020)Goldstein, Fletcher, Veres, Briggs,
      Cleveland, Gibby, Hui, Bissaldi, Burns, Hamburg, \& et~al.}]{goldstein_2020}
    Goldstein, A., Fletcher, C., Veres, P., {et~al.} 2020, The Astrophysical
      Journal, 895, 40, \dodoi{10.3847/1538-4357/ab8bdb}
    
    \bibitem[{Groom(2004)}]{groom_2004}
    Groom, D. 2004, Experimental Astronomy, 14, 81, \dodoi{10.1007/1-4020-2527-0_9}
    
    \bibitem[{{Howard} {et~al.}(2019){Howard}, {Corbett}, {Law}, {Ratzloff},
      {Glazier}, {Fors}, {del Ser}, \& {Haislip}}]{evryflare_i}
    {Howard}, W.~S., {Corbett}, H., {Law}, N.~M., {et~al.} 2019, \apj, 881, 9,
      \dodoi{10.3847/1538-4357/ab2767}
    
    \bibitem[{{Howard} {et~al.}(2018){Howard}, {Tilley}, {Corbett}, {Youngblood},
      {Loyd}, {Ratzloff}, {Law}, {Fors}, {del Ser}, {Shkolnik}, {Ziegler}, {Goeke},
      {Pietraallo}, \& {Haislip}}]{proxima_superflare}
    {Howard}, W.~S., {Tilley}, M.~A., {Corbett}, H., {et~al.} 2018, \apjl, 860,
      L30, \dodoi{10.3847/2041-8213/aacaf3}
    
    \bibitem[{{IceCube Collaboration} {et~al.}(2017){IceCube Collaboration},
      Aartsen, Ackermann, Adams, Aguilar, Ahlers, Ahrens, Altmann, Andeen,
      Anderson, \& et~al.}]{icecube_instrument}
    {IceCube Collaboration}, Aartsen, M.~G., Ackermann, M., {et~al.} 2017, Journal
      of Instrumentation, 12, P03012–P03012,
      \dodoi{10.1088/1748-0221/12/03/p03012}
    
    \bibitem[{Johnson(1949)}]{johnson_1949}
    Johnson, N.~L. 1949, Biometrika, 36, 149, \dodoi{10.1093/biomet/36.1-2.149}
    
    \bibitem[{{Karpov} {et~al.}(2016){Karpov}, {Beskin}, {Biryukov}, {Bondar},
      {Ivanov}, {Katkova}, {Perkov}, \& {Sasyuk}}]{karpov_2016}
    {Karpov}, S., {Beskin}, G., {Biryukov}, A., {et~al.} 2016, in Revista Mexicana
      de Astronomia y Astrofisica Conference Series, Vol.~48, Revista Mexicana de
      Astronomia y Astrofisica Conference Series, 91--96
    
    \bibitem[{{Law} {et~al.}(2015){Law}, {Fors}, {Ratzloff}, {Wulfken},
      {Kavanaugh}, {Sitar}, {Pruett}, {Birchard}, {Barlow}, {Cannon}, {Cenko},
      {Dunlap}, {Kraus}, \& {Maccarone}}]{evryscope_project_paper}
    {Law}, N.~M., {Fors}, O., {Ratzloff}, J., {et~al.} 2015, \pasp, 127, 234,
      \dodoi{10.1086/680521}
    
    \bibitem[{{LeCun} {et~al.}(1998){LeCun}, {Bottou}, {Bengio}, \&
      {Haffner}}]{lecun_convnets}
    {LeCun}, Y., {Bottou}, L., {Bengio}, Y., \& {Haffner}, P. 1998, Proceedings of
      the IEEE, 86, no. 11, 2278
    
    \bibitem[{{Lipunov} {et~al.}(2004){Lipunov}, {Krylov}, {Kornilov}, {Borisov},
      {Kuvshinov}, {Belinsky}, {Kuznetsov}, {Potanin}, {Antipov}, {Tyurina},
      {Gorbovskoy}, \& {Chilingaryan}}]{master_instrument}
    {Lipunov}, V.~M., {Krylov}, A.~V., {Kornilov}, V.~G., {et~al.} 2004,
      Astronomische Nachrichten, 325, 580, \dodoi{10.1002/asna.200410284}
    
    \bibitem[{{LSST Science Collaboration} {et~al.}(2009){LSST Science
      Collaboration}, {Abell}, {Allison}, {Anderson}, {Andrew}, {Angel}, {Armus},
      {Arnett}, {Asztalos}, {Axelrod}, {Bailey}, {Ballantyne}, {Bankert},
      {Barkhouse}, {Barr}, {Barrientos}, {Barth}, {Bartlett}, {Becker}, {Becla},
      {Beers}, {Bernstein}, {Biswas}, {Blanton}, {Bloom}, {Bochanski}, {Boeshaar},
      {Borne}, {Bradac}, {Brandt}, {Bridge}, {Brown}, {Brunner}, {Bullock},
      {Burgasser}, {Burge}, {Burke}, {Cargile}, {Chand rasekharan}, {Chartas},
      {Chesley}, {Chu}, {Cinabro}, {Claire}, {Claver}, {Clowe}, {Connolly}, {Cook},
      {Cooke}, {Cooray}, {Covey}, {Culliton}, {de Jong}, {de Vries}, {Debattista},
      {Delgado}, {Dell'Antonio}, {Dhital}, {Di Stefano}, {Dickinson}, {Dilday},
      {Djorgovski}, {Dobler}, {Donalek}, {Dubois-Felsmann}, {Durech},
      {Eliasdottir}, {Eracleous}, {Eyer}, {Falco}, {Fan}, {Fassnacht}, {Ferguson},
      {Fernandez}, {Fields}, {Finkbeiner}, {Figueroa}, {Fox}, {Francke}, {Frank},
      {Frieman}, {Fromenteau}, {Furqan}, {Galaz}, {Gal-Yam}, {Garnavich},
      {Gawiser}, {Geary}, {Gee}, {Gibson}, {Gilmore}, {Grace}, {Green}, {Gressler},
      {Grillmair}, {Habib}, {Haggerty}, {Hamuy}, {Harris}, {Hawley}, {Heavens},
      {Hebb}, {Henry}, {Hileman}, {Hilton}, {Hoadley}, {Holberg}, {Holman},
      {Howell}, {Infante}, {Ivezic}, {Jacoby}, {Jain}, {R}, {Jedicke}, {Jee},
      {Garrett Jernigan}, {Jha}, {Johnston}, {Jones}, {Juric}, {Kaasalainen},
      {Styliani}, {Kafka}, {Kahn}, {Kaib}, {Kalirai}, {Kantor}, {Kasliwal},
      {Keeton}, {Kessler}, {Knezevic}, {Kowalski}, {Krabbendam}, {Krughoff},
      {Kulkarni}, {Kuhlman}, {Lacy}, {Lepine}, {Liang}, {Lien}, {Lira}, {Long},
      {Lorenz}, {Lotz}, {Lupton}, {Lutz}, {Macri}, {Mahabal}, {Mandelbaum},
      {Marshall}, {May}, {McGehee}, {Meadows}, {Meert}, {Milani}, {Miller},
      {Miller}, {Mills}, {Minniti}, {Monet}, {Mukadam}, {Nakar}, {Neill}, {Newman},
      {Nikolaev}, {Nordby}, {O'Connor}, {Oguri}, {Oliver}, {Olivier}, {Olsen},
      {Olsen}, {Olszewski}, {Oluseyi}, {Padilla}, {Parker}, {Pepper}, {Peterson},
      {Petry}, {Pinto}, {Pizagno}, {Popescu}, {Prsa}, {Radcka}, {Raddick},
      {Rasmussen}, {Rau}, {Rho}, {Rhoads}, {Richards}, {Ridgway}, {Robertson},
      {Roskar}, {Saha}, {Sarajedini}, {Scannapieco}, {Schalk}, {Schindler},
      {Schmidt}, {Schmidt}, {Schneider}, {Schumacher}, {Scranton}, {Sebag},
      {Seppala}, {Shemmer}, {Simon}, {Sivertz}, {Smith}, {Allyn Smith}, {Smith},
      {Spitz}, {Stanford}, {Stassun}, {Strader}, {Strauss}, {Stubbs}, {Sweeney},
      {Szalay}, {Szkody}, {Takada}, {Thorman}, {Trilling}, {Trimble}, {Tyson}, {Van
      Berg}, {Vand en Berk}, {VanderPlas}, {Verde}, {Vrsnak}, {Walkowicz}, {Wand
      elt}, {Wang}, {Wang}, {Warner}, {Wechsler}, {West}, {Wiecha}, {Williams},
      {Willman}, {Wittman}, {Wolff}, {Wood-Vasey}, {Wozniak}, {Young}, {Zentner},
      \& {Zhan}}]{LSST_Science_Book}
    {LSST Science Collaboration}, {Abell}, P.~A., {Allison}, J., {et~al.} 2009,
      arXiv e-prints, arXiv:0912.0201.
    \newblock \doarXiv{0912.0201}
    
    \bibitem[{Lyutikov \& Lorimer(2016)}]{lyutikov_2016}
    Lyutikov, M., \& Lorimer, D.~R. 2016, The Astrophysical Journal Letters, 824,
      L18, \dodoi{10.3847/2041-8205/824/2/l18}
    
    \bibitem[{Maley(1991)}]{Maley_1991}
    Maley, P. 1991, Advances in Space Research, 11, 33–36,
      \dodoi{10.1016/0273-1177(91)90539-v}
    
    \bibitem[{Maley(1987)}]{Maley_1987}
    Maley, P.~D. 1987, The Astrophysical Journal, 317, L39, \dodoi{10.1086/184909}
    
    \bibitem[{{McDowell}(2020)}]{mcdowell_2020}
    {McDowell}, J.~C. 2020, \apjl, 892, L36, \dodoi{10.3847/2041-8213/ab8016}
    
    \bibitem[{Rast(1991)}]{Rast_1991}
    Rast, R.~H. 1991, Icarus, 90, 328–329, \dodoi{10.1016/0019-1035(91)90112-7}
    
    \bibitem[{{Ratzloff} {et~al.}(2020){Ratzloff}, {Law}, {Corbett}, {Fors}, \&
      {del Ser}}]{ratzloff_robotilter}
    {Ratzloff}, J.~K., {Law}, N.~M., {Corbett}, H.~T., {Fors}, O., \& {del Ser}, D.
      2020, arXiv e-prints, arXiv:2001.00879.
    \newblock \doarXiv{2001.00879}
    
    \bibitem[{{Ratzloff} {et~al.}(2019){Ratzloff}, {Law}, {Fors}, {Corbett},
      {Howard}, {del Ser}, \& {Haislip}}]{evryscope_instrument_paper}
    {Ratzloff}, J.~K., {Law}, N.~M., {Fors}, O., {et~al.} 2019, \pasp, 131, 075001,
      \dodoi{10.1088/1538-3873/ab19d0}
    
    \bibitem[{{Schaefer} {et~al.}(1987){Schaefer}, {Barber}, {Brooks}, {Deforrest},
      {Maley}, {McLeod}, {McNiel}, {Noymer}, {Presnell}, {Schwartz}, \&
      {Whitney}}]{schaefer_1987}
    {Schaefer}, B.~E., {Barber}, M., {Brooks}, J.~J., {et~al.} 1987, \apj, 320,
      398, \dodoi{10.1086/165552}
    
    \bibitem[{{Shamir} \& {Nemiroff}(2006)}]{shamir_2006}
    {Shamir}, L., \& {Nemiroff}, R.~J. 2006, \pasp, 118, 1180,
      \dodoi{10.1086/506989}
    
    \bibitem[{{Shappee} {et~al.}(2014){Shappee}, {Prieto}, {Grupe}, {Kochanek},
      {Stanek}, {De Rosa}, {Mathur}, {Zu}, {Peterson}, {Pogge}, {Komossa}, {Im},
      {Jencson}, {Holoien}, {Basu}, {Beacom}, {Szczygie{\l}}, {Brimacombe},
      {Adams}, {Campillay}, {Choi}, {Contreras}, {Dietrich}, {Dubberley},
      {Elphick}, {Foale}, {Giustini}, {Gonzalez}, {Hawkins}, {Howell}, {Hsiao},
      {Koss}, {Leighly}, {Morrell}, {Mudd}, {Mullins}, {Nugent}, {Parrent},
      {Phillips}, {Pojmanski}, {Rosing}, {Ross}, {Sand}, {Terndrup}, {Valenti},
      {Walker}, \& {Yoon}}]{asassn_instrument}
    {Shappee}, B.~J., {Prieto}, J.~L., {Grupe}, D., {et~al.} 2014, \apj, 788, 48,
      \dodoi{10.1088/0004-637X/788/1/48}
    
    \bibitem[{{Talens} {et~al.}(2017){Talens}, {Spronck}, {Lesage}, {Otten},
      {Stuik}, {Pollacco}, \& {Snellen}}]{talens_2017}
    {Talens}, G.~J.~J., {Spronck}, J.~F.~P., {Lesage}, A.~L., {et~al.} 2017, \aap,
      601, A11, \dodoi{10.1051/0004-6361/201630319}
    
    \bibitem[{{The CHIME/FRB Collaboration} {et~al.}(2018){The CHIME/FRB
      Collaboration}, :, Amiri, Bandura, Berger, Bhardwaj, Boyce, Boyle, Brar,
      Burhanpurkar, \& et~al.}]{chime_2018}
    {The CHIME/FRB Collaboration}, :, Amiri, M., {et~al.} 2018, The Astrophysical
      Journal, 863, 48, \dodoi{10.3847/1538-4357/aad188}
    
    \bibitem[{{Tingay}(2020)}]{tingay_2020}
    {Tingay}, S. 2020, \pasa, 37, e015, \dodoi{10.1017/pasa.2020.7}
    
    \bibitem[{Tonry {et~al.}(2018)Tonry, Denneau, Heinze, Stalder, Smith, Smartt,
      Stubbs, Weiland, \& Rest}]{atlas_instrument}
    Tonry, J.~L., Denneau, L., Heinze, A.~N., {et~al.} 2018, Publications of the
      Astronomical Society of the Pacific, 130, 064505,
      \dodoi{10.1088/1538-3873/aabadf}
    
    \bibitem[{{Tonry} {et~al.}(2018){Tonry}, {Denneau}, {Flewelling}, {Heinze},
      {Onken}, {Smartt}, {Stalder}, {Weiland}, \& {Wolf}}]{atlas_catalog}
    {Tonry}, J.~L., {Denneau}, L., {Flewelling}, H., {et~al.} 2018, \apj, 867, 105,
      \dodoi{10.3847/1538-4357/aae386}
    
    \bibitem[{{van Dokkum}(2001)}]{lacosmic}
    {van Dokkum}, P.~G. 2001, \pasp, 113, 1420, \dodoi{10.1086/323894}
    
    \bibitem[{Yang {et~al.}(2019)Yang, Zhang, \& Wei}]{yang_2019}
    Yang, Y.-P., Zhang, B., \& Wei, J.-Y. 2019, The Astrophysical Journal, 878, 89,
      \dodoi{10.3847/1538-4357/ab1fe2}
    
    \bibitem[{Yeh {et~al.}(1996)Yeh, Lee, \& Kremers}]{yeh_lee_kremers_1996}
    Yeh, T., Lee, B.~B., \& Kremers, J. 1996, Vision Research, 36, 913–931,
      \dodoi{10.1016/0042-6989(95)00332-0}
    
    \bibitem[{{Zackay} {et~al.}(2016){Zackay}, {Ofek}, \& {Gal-Yam}}]{zogy}
    {Zackay}, B., {Ofek}, E.~O., \& {Gal-Yam}, A. 2016, \apj, 830, 27,
      \dodoi{10.3847/0004-637X/830/1/27}
    
\end{thebibliography}

\bibliographystyle{aasjournal}

\end{document}